\documentclass[12pt]{article}
\usepackage{axodraw,epsfig}

\newcommand{\inv}{\left. g^0_A \right|_{\rm inv}}

\newcommand{\Frac}[2]%
{{\textstyle \frac{\mbox{\footnotesize $#1$}\rule[-0.9mm]{0mm}{1mm}}%
{\mbox{\footnotesize $#2$}\rule{0mm}{3.1mm}}}}
\renewcommand{\thefootnote}{\fnsymbol{footnote}}

\begin{document}
\begin{titlepage}
\vspace*{-12 mm}
\noindent
\begin{flushright}
\begin{tabular}{l@{}}
\end{tabular}
\end{flushright}
\vskip 12 mm
\begin{center}
{\Large \bf
Final state interaction \\
\vspace{2ex}
and a light mass ``exotic'' resonance
} 
\\[10mm]
{\bf Steven D. Bass$^{[a]}$ and Eugenio Marco$^{[b]}$}
\\[10mm]   
{\em 
$^a$ 
ECT*, Strada delle Tabarelle 286, I-38050 Villazzano, Trento, Italy
 } \\
\vspace{3ex}
\vspace{3ex}
{\em 
$^b$ 
Physik Department, Technische Universit\"at M\"unchen, \\
D-85747 Garching, Germany } 
\end{center}
\vskip 10 mm
\begin{abstract}
\noindent
We investigate the possibility that the light-mass exotic mesons with 
$J^{PC} = 1^{-+}$ observed at BNL and CERN may be resonances
in the $\eta \pi$ and $\eta' \pi$ systems associated with the 
anomalous glue which generates the $\eta'$ mass in QCD.

\end{abstract}

\vspace{2.0cm}
\begin{flushleft}
\end{flushleft} 
\end{titlepage}
\renewcommand{\labelenumi}{(\alph{enumi})}
\renewcommand{\labelenumii}{(\roman{enumii})}
\renewcommand{\thefootnote}{\arabic{footnote}}
\newpage

\section{Introduction}

Recent experiments at BNL \cite{e852a,e852b} and CERN \cite{crystalb}
have revealed evidence for QCD ``exotic'' meson states with quantum 
numbers $J^{PC}=1^{-+}$.
These mesons are particularly interesting because the quantum numbers 
$J^{PC}=1^{-+}$ are inconsistent with a simple quark-antiquark bound 
state suggesting a possible ``valence'' gluonic component 
-- for example through coupling to the operator
$[ {\bar q} \gamma_{\mu} q G^{\mu \nu}$ ].
Two such exotics, denoted $\pi_1$, have been observed through
$\pi^- p \rightarrow \pi_1 p$ at BNL:
with masses 1400 MeV (in decays to $\eta \pi$) 
\cite{e852a}
and 1600 MeV (in decays to $\eta' \pi$ and $\rho \pi$) \cite{e852b}.
The $\pi_1 (1400)$ state has also been observed in ${\bar p} N$
processes by the Crystal Barrel Collaboration at CERN \cite{crystalb}. 
These states are considerably lighter than the predictions (about 
1900 MeV) of quenched lattice QCD \cite{lattice}
and QCD inspired models \cite{models} for the 
lowest mass $q {\bar q} g$ state with quantum numbers $J^{PC}=1^{-+}$.
While chiral corrections may bring the lattice predictions down by 
about 100 MeV \cite{thomas} these results suggest that, perhaps, 
the ``exotic'' states observed by the experimentalists might involve 
significant meson-meson bound state contributions.
Furthermore, the decays to $\eta$ or $\eta'$ mesons plus a pion may 
hint at a possible connection to axial U(1) dynamics.

In this paper we investigate the possibility that the observed exotics 
may be resonances in $\eta - \pi$ and $\eta' - \pi$ scattering,
possibly induced by the non-perturbative glue which generates
the $\eta'$ mass -- the famous axial U(1) problem of QCD \cite{zuozu1}.
Unitarisation methods for chiral Lagrangians have shown
that it is possible to dynamically generate resonances in the
meson-baryon and meson-meson sectors \cite{wwa,OseRam,OllOse97,OllPel}.
In particular,
in the meson-meson sector resonances like the $\sigma, f_0, a_0$
which decay into two particles have been succesfully generated
using coupled channels and the Bethe-Salpeter equation.

We investigate the $\eta \pi$ and $\eta' \pi$ systems through the 
Bethe-Salpeter equation in a coupled channels approach with 
potentials derived from the leading order, $O(p^2)$, axial U(1) 
extended chiral Lagrangian.
Before we begin our coupled channels discussion first consider 
the scenario that the observed processes involved production of 
a simple ``hybrid''
meson with valence $q {\bar q} g$ wavefunction which decays to
$\eta \pi$ and $\eta' \pi$ through flavour-singlet (OZI-violating)
gluonic coupling of the ``valence gluon'' to the $\eta$ and $\eta'$.
In Born approximation (no final state interaction)
one would predict
\begin{equation}
r_{\pi_1} = 
 { {\Gamma}(\pi_1 \rightarrow \eta' \pi) 
   \over {\Gamma}(\pi_1 \rightarrow \eta  \pi) } 
= 
\cot^2 \theta 
\sqrt{ { (M_{\pi_1}^2 - (m_{\eta'} + m_{\pi})^2 ) 
         (M_{\pi_1}^2 - (m_{\eta'} - m_{\pi})^2 )
\over
         (M_{\pi_1}^2 - (m_{\eta} + m_{\pi})^2 ) 
         (M_{\pi_1}^2 - (m_{\eta} - m_{\pi})^2 ) }}
\end{equation}
where $\theta$ is the $\eta - \eta'$ mixing angle.
Taking
$\theta = -18$ 
degrees 
\cite{gilman,frere}
Eq.(1) evaluates to
$r_{\rm 1400} =
 {\Gamma}(\pi_1(1400) \rightarrow \eta' \pi)/
 {\Gamma}(\pi_1(1400) \rightarrow \eta  \pi)$ = 5.7
and 
$r_{\rm 1600} =
 {\Gamma}(\pi_1(1600) \rightarrow \eta' \pi)/
 {\Gamma}(\pi_1(1600) \rightarrow \eta  \pi) = 6.8$
for the 1400 and 1600 MeV states.
The absence of an observed signal for 
$\pi_1 (1400) \rightarrow \eta' \pi$
and 
$\pi_1 (1600) \rightarrow \eta  \pi$
suggests that this scenario is too simple and that final state
interaction and/or other hadronic physics is important to understanding
the observed exotics.

We now discuss the chiral Lagrangian in Section 2 and use this in Section 3 
within the Bethe-Salpeter formalism to investigate possible resonance 
structures with exotic quantum numbers.

\section{The low-energy effective Lagrangian}

We start from the low-energy effective Lagrangian \cite{vecca,lagran} 
\begin{eqnarray}\label{eq:L1}
{\cal L}_{\rm m} = & &
{F_{\pi}^2 \over 4} 
{\rm Tr}(\partial^{\mu}U \partial_{\mu}U^{\dagger}) 
+
{F_{\pi}^2 \over 4} {\rm Tr} \chi_0 \ \biggl( U + U^{\dagger} \biggr)
\\ \nonumber
&+& 
  {1 \over 2} i Q {\rm Tr} \biggl[ \log U - \log U^{\dagger} \biggr]
+ {3 \over {\tilde m}_{\eta_0}^2 F_{0}^2} Q^2.
\end{eqnarray}
where $U$ is the unitary meson matrix 
\begin{equation}
U = \exp \ \biggl(  i {\phi \over F_{\pi}}  
                  + i \sqrt{2 \over 3} {\eta_0 \over F_0} \biggr) .
\end{equation}
Here 
\begin{equation}
\phi = \ \sqrt{2}
\left(\begin{array}{ccc} 
{1 \over \sqrt{2}} \pi^0 + {1 \over \sqrt{6}} \eta_8 & \pi^+ & K^+ \\
\\
\pi^- & 
-{1 \over \sqrt{2}} \pi^0 + {1\over \sqrt{6}} \eta_8 & K^0 \\
\\
K^- & {\bar K}^0 & -{2 \over \sqrt{6}} \eta_8 
\vphantom{\inv}  
\end{array}\right) 
\end{equation}
denotes the octet of would-be Goldstone bosons associated with 
spontaneous chiral $SU(3)_L \otimes SU(3)_R$ breaking, 
$\eta_0$ 
is the singlet boson and $Q$
denotes the topological charge density
$Q = {\alpha_s \over 4 \pi} G_{\mu \nu} {\tilde G}^{\mu \nu}$
.
The pion decay constant $F_{\pi} = 92.4$MeV;
$F_0$ renormalises the flavour-singlet decay constant 
and
$\chi_0 = 
{\rm diag} [ m_{\pi}^2, m_{\pi}^2, (2 m_K^2 - m_{\pi}^2 ) ]$
is the meson mass matrix.

The $U_A(1)$ gluonic potential involving the topological charge 
density is introduced to generate the gluonic contribution 
to the $\eta$ and $\eta'$ masses and 
to reproduce the anomaly \cite{adler,bell} in the divergence of 
the gauge-invariantly renormalised flavour-singlet axial-vector 
current,
viz.
\begin{equation}
\partial^\mu J_{\mu5} =
\sum_{k=1}^{f} 2 i \biggl[ m_k \bar{q}_k \gamma_5 q_k
\biggr]_{GI}
+ N_f
\biggl[ {\alpha_s \over 4 \pi} G_{\mu \nu} {\tilde G}^{\mu \nu}
\biggr]_{GI}^{\mu^2}
\end{equation}
where
\begin{equation}
J_{\mu 5} =
\left[ \bar{u}\gamma_\mu\gamma_5u
                  + \bar{d}\gamma_\mu\gamma_5d
                  + \bar{s}\gamma_\mu\gamma_5s \right]_{GI}^{\mu^2} .
\end{equation} 
The gluonic term $Q$ is treated as a background field with no kinetic 
term.
It may be eliminated through its equation of motion to generate a 
gluonic mass term for the singlet boson,
viz.
\begin{equation}
{1 \over 2} i Q {\rm Tr} \biggl[ \log U - \log U^{\dagger} \biggr]
+ {3 \over {\tilde m}_{\eta_0}^2 F_{0}^2} Q^2 
\
\mapsto \
- {1 \over 2} {\tilde m}_{\eta_0}^2 \eta_0^2 
\end{equation}
Coupling the Lagrangian (2) to the baryon octet allows one to study
the $\eta$-nucleon and $\eta'$-nucleon interactions \cite{sb99,bww}
where one finds a gluonic induced contact term \cite{sb99} in the 
low-energy
$pp \rightarrow pp \eta$ and
$pp \rightarrow pp \eta'$ reactions, 
which are presently under vigorous experimental 
study at CELSIUS \cite{celsius} and COSY \cite{cosy}.

In this paper we shall be interested in meson scattering processes involving 
$\eta' \pi \leftrightarrow (\eta' \pi, \eta \pi)$.
Fourth-order terms in the meson fields are induced by the first two terms in 
Eq.(2) and also by the $U_A(1)$ invariant term 
\cite{veccb}
\begin{equation}\label{eq:L2}
{\cal L}_{m2Q} \ = \
\lambda \ Q^2 \ {\rm Tr} \ \partial_{\mu} U \partial^{\mu} U^{\dagger}
\end{equation}
which is important in describing the $\eta \eta' \pi \pi$ system.
After we eliminate $Q$ through its equation of motion
the $Q$ dependent part of the effective Lagrangian $(2)+(8)$
becomes
\begin{eqnarray}
{\cal L}_Q 
&=&
- {1 \over 2} \ {\tilde m}_{\eta_0}^2 \eta_0^2
  + {1 \over 6} \ \lambda \ {\tilde m}_{\eta_0}^4 \eta_0^2 \ F_0^2 \
  {\rm Tr} \ \partial_{\mu} U \partial^{\mu} U^{\dagger} + ... \\ \nonumber
&=&
- {1 \over 2} {\tilde m}_{\eta_0}^2 \ \eta_0^2 \ 
+
{3 \over 2} \ \eta_0^2 \ 
\beta 
\ \biggl( {1 \over F_{\pi}} \biggr)^2 \ 
\partial^{\mu} \pi_a \partial_{\mu} \pi_a 
+
{3 \over 2} \ \eta_0^2 \
\ \beta \
\ \biggl( {1 \over F_0} \biggr)^2 \ 
\partial_{\mu} \eta_0 \partial^{\mu} \eta_0 \ + ...
\end{eqnarray}
where
we have set $\lambda \equiv {9 \over 2 F_0^2 {\tilde m}_{\eta_0}^4} \beta$.
In general, one can expect OZI violation wherever a coupling involving 
the $Q$-field occurs.
The value of $F_0$ is usually determined from the decay rate for 
$\eta' \rightarrow 2 \gamma$.
In QCD one finds the relation \cite{shore}
\begin{equation}
{2 \alpha \over \pi} = 
\sqrt{3 \over 2} F_0 \biggl( g_{\eta' \gamma \gamma} - g_{Q \gamma \gamma} 
\biggr) 
\end{equation}
The observed decay rate \cite{twogamma} is consistent \cite{gilman} 
with the OZI prediction for $g_{\eta' \gamma \gamma}$ 
{\it if} $F_0$ and 
$g_{Q \gamma \gamma}$ take their OZI values: 
$F_0 \simeq F_{\pi}$ and $g_{Q \gamma \gamma} = 0$.
In the rest of this paper we set $F_0 = F_{\pi}$.

\section{Resonances in $\eta \pi$ and $\eta' \pi$ scattering}

From the Lagrangian of Eqs.\ (\ref{eq:L1}) and (\ref{eq:L2}) 
one obtains the $T$ matrix elements at tree level, which, in 
this paper, we will call potentials $V$. 
We label the channels: 
$1=K \bar{K}$, $2=\pi \eta_8$, $3=\pi \eta_0$.
The potentials involving the $\pi \eta_8$ and $K {\bar K}$
channels are derived from the SU(3) part of the Lagrangian
and are given in Ref.\ \cite{OllOse97}.
We write below the three additional combinations involving
the singlet boson that we will need in our coupled channels
analysis:
\begin{equation}
V_{13} = \frac{\sqrt{3}}{9 f_{\pi}^2} (2 m_K^2 + m_{\pi}^2)\, ,
\end{equation}
\begin{equation}
V_{23} =  \frac{-\sqrt{2}}{3 f_{\pi}^2}  m_{\pi}^2\, ,
\end{equation}
and
\begin{equation}\label{eq:v33}
V_{33} =  \frac{-2}{3 f_{\pi}^2}  m_{\pi}^2
+ \frac{3\beta}{f_{\pi}^2} (t - 2 m_{\pi}^2)
\end{equation}
We work in the one mixing angle scheme
\begin{eqnarray}
| \eta \rangle &=& 
\cos \theta \ | \eta_8 \rangle - \sin \theta \ | \eta_0 \rangle
\\ \nonumber
| \eta' \rangle &=& 
\sin \theta \ | \eta_8 \rangle + \cos \theta \ | \eta_0 \rangle
\end{eqnarray}
where the mixing angle $\theta$ is taken as -18 degrees 
\cite{gilman,frere}.
We refer the reader to \cite{leutwyler,feldmann} for a discussion of 
the two-mixing angle scheme which enters if the chiral Lagrangian is
extended to $O(p^4)$ in the meson momenta.

In order to generate the resonances we follow the method developed
in Ref.\ \cite{OllOse97}, where the $a_0$, $\sigma$ and $f_0$ are
generated dynamically using coupled channels and the Bethe-Salpeter
equation. 
One can write the $T$ matrix as
\begin{equation}
T_{ij} = V_{ij} + \sum_k V_{ik} G_k T_{kj}\, ,
\end{equation}
and then extract $T$, 
which in matrix form can be written as $T=(1-VG)^{-1}V$ where $G_k$ 
represents a loop with the mesons of channel $k$.
The meson loop is regularised with a three-momentum cut-off 
which, following \cite{OllOse97,phidecay}, 
we take as $\Lambda = 900$ MeV for all three channels.
This three-momentum momentum cut-off is the only model parameter in 
our calculation and its value is fixed in order to reproduce the
$a_0(980)$ mass and width. A change of $\pm 100$ MeV in the value of 
$\Lambda$ induces a change of $\pm 30$ MeV for the $a_0(980)$ pole 
mass and no variation of its width.

We investigate possible $s$ and $p$-wave resonances via the projection
\begin{equation}
T_L = \frac{1}{2}\int_{-1}^1 d\cos\theta P_L(\cos\theta) T .
\end{equation}
For $s$-waves all three channels are involved, whereas for $p$-waves
there is a decoupling of the channels, since they have different
quantum numbers. 
$K\bar{K}$ has $J^{PC}=1^{--}$, 
whereas $\pi \eta$ and $\pi \eta'$ are in $J^{PC}=1^{-+}$
-- the quantum numbers of the exotic. 
Here we consider only $p$ waves in the $J^{PC}=1^{-+}$ channel.

The parameter $\beta$ can be determined via the $\eta'$ decay 
into $\pi \pi \eta$, 
\begin{equation}
T_{\eta ' \rightarrow \eta \pi^0 \pi^0} =
\frac{1}{2} \sin{2 \theta} \ 
\biggl(
T_{\eta_8 \rightarrow \eta_8 \pi^0 \pi^0}
- T_{\eta_0 \rightarrow \eta_0 \pi^0 \pi^0} \biggr)
\ + \ \cos 2 \theta \ T_{\eta_0 \rightarrow \eta_8 \pi^0 \pi^0}
\end{equation}
where 
$
T_{\eta_8 \rightarrow \eta_8 \pi^0 \pi^0} = \frac{-m_{\pi}^2}{3 f_{\pi}^2}
$ \cite{OllOse97},
$T_{\eta_0 \rightarrow \eta_8 \pi^0 \pi^0} = V_{23}$ and
$T_{\eta_0 \rightarrow \eta_0 \pi^0 \pi^0} = V_{33}$.
One finds
two possible solutions: $\beta = -0.625$ and $\beta = 0.86$. 
We can investigate the resonance structures which result 
from each of these two possibilities and try to extract consequences.

With $\beta=-0.625$ a resonance structure appears in the $p$-wave
amplitude. In Fig.\ 1 we show the amplitude squared for the different
channels involved. One can observe a resonance structure at around
1400 MeV, most prominent in the $\pi \eta' \rightarrow \pi \eta'$
channel. By looking at the second Riemann sheet on the complex $s$ plane 
the pole can be found at $(1399,-153)$ MeV, 
a result remarkably close to the experimentally 
determined values for the $\pi_1(1400)$: 
mass $1370 \pm 16$ MeV and width $385 \pm 40$ MeV.
With this value of $\beta$, there is no $s$-wave pole around 1450 MeV.
\begin{figure}[H]
\centerline{
\includegraphics[width=0.6\textwidth,angle=-90]{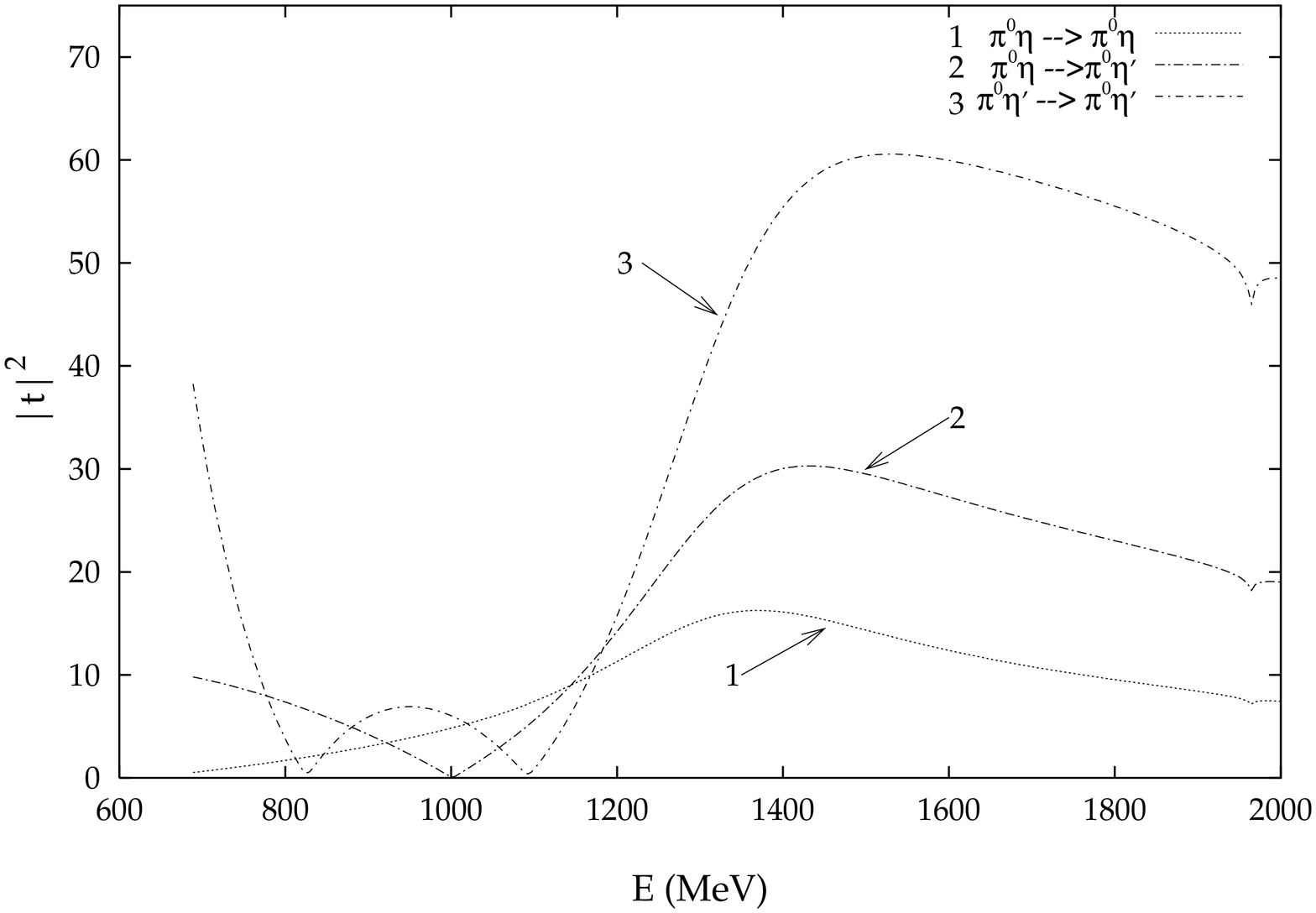}
}
\caption{}
\label{fig2}
\end{figure}
Since our exotic is built exclusively through the $p$-wave
part of the $\pi \eta_0 \rightarrow \pi \eta_0$ amplitude,
it decays mainly (86\%) into $\pi \eta'$ and 14\% to $\pi \eta$. 
These values clearly indicate that one would need to include 
higher order terms to get the branching ratios correct. 
However, 
these new terms would not change the position of the exotic, 
which is determined by the term $(1-VG)^{-1}$ with a $\pi \eta'$ 
in the loop. 
The new terms involving higher order $\pi \eta_8$ contributions 
could act to increase the amplitude for 
$\pi \eta_8 \rightarrow \pi \eta_8$ while keeping the same resonance structure.

The opposite situation is found in the second scenario, taking
$\beta=0.86$. 
Here we find an s-wave resonance structure around 1300 MeV, see
Fig.\ 2, where all the different squared amplitudes are shown.
Looking at the complex $s$ plane we find the pole at 
$(1286,-93)MeV$.
This resonance structure is a candidate for the scalar $a_0(1450)$
which has measured width $265 \pm 13$MeV \cite{twogamma}.
The branching ratios are:
39\% into channel 1, 40\% into channel 2 and 21\% into channel 3.
No resonance structure appears in the $p$-wave amplitude with 
$\beta =0.86$.

These results are essentially insensitive to variations in the loop-momentum
cut-off $\Lambda$.
For the scalar $a_0(1450)$ candidate varying the cut-off parameter 
$\Lambda$ by $\pm 100$ MeV changes the mass of the scalar resonance
by approximately $\pm 10$ MeV. 
Decreasing $\Lambda$ to 800 MeV increases the width by about 40 MeV
while increasing $\Lambda$ to 1 GeV decreases the width by just 2MeV.
For the resonance in the $p$-wave amplitude a change in 
$\Lambda$ by $\pm 100$MeV induces no variation of the resonance
mass, while its width changes by $\pm 100$ MeV.

One can easily understand why different resonances are generated
with the different signs of $\beta$. In order to have attraction the
potential has to have a negative sign given our conventions. Once
the $s$-wave is projected in $V_{33}$, there are two terms in it,
a negative term which depends on the masses
(see Eq.\ (\ref{eq:v33})) and the $\beta$ term
with a minus sign in front of it. A positive $\beta$ adds attraction
and the scalar resonance is generated. On the contrary a negative
sign generates repulsion and the resonance is not generated. The
$p$-wave part has the opposite behaviour. It has a $\beta$ term with
a positive sign and thus only a negative value of $\beta$ generates
the required attraction to generate the resonance.

\begin{figure}[H]
\centerline{
\includegraphics[width=0.6\textwidth,angle=-90]{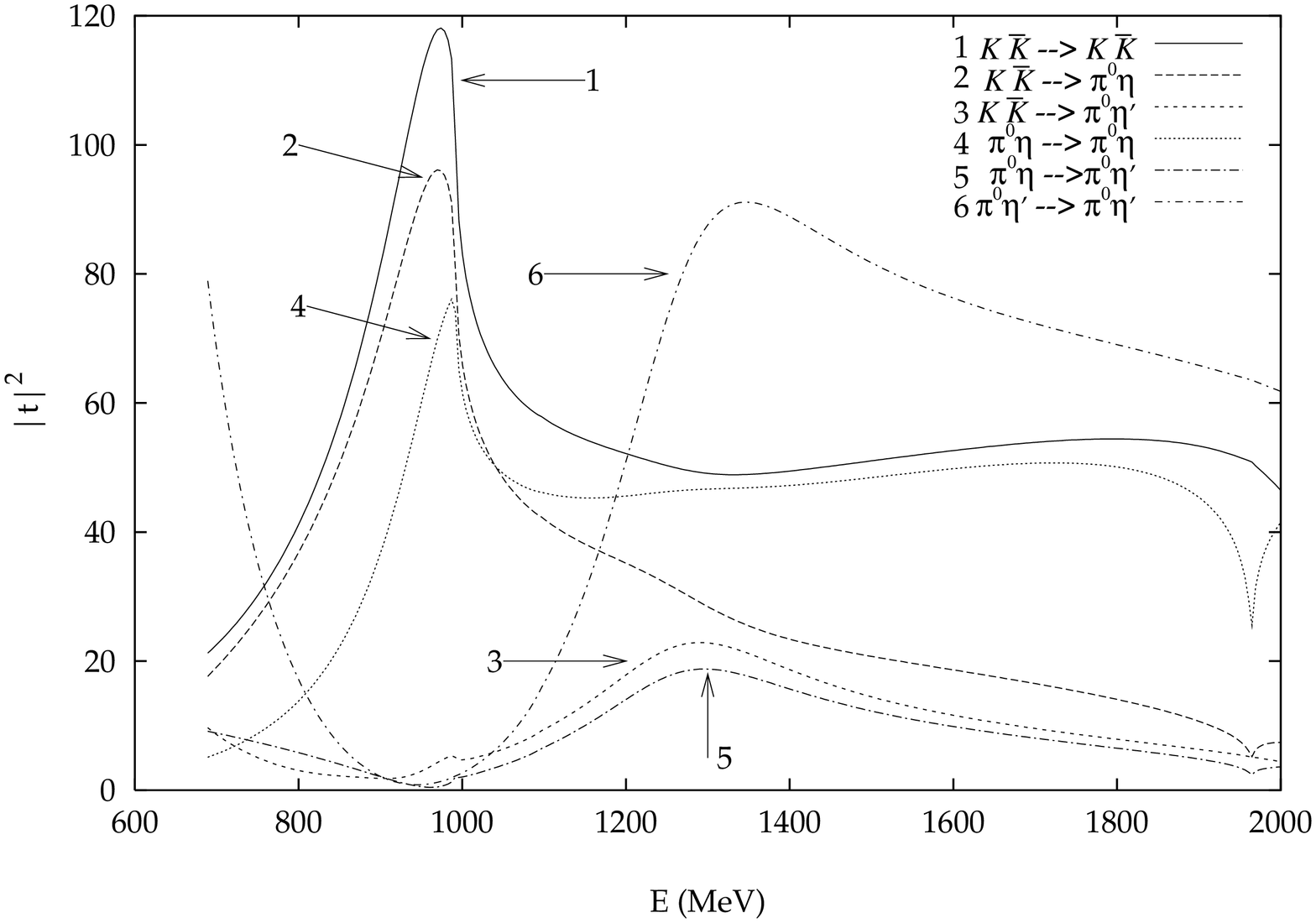}
}
\caption{}
\label{fig1}
\end{figure}

Although we have no way of discarding any of the solutions and possibly 
higher order terms would be needed to have a more complete picture, it 
seems clear that this simple calculation illustrates the possibility to 
describe the
appearance of exotics by means of a coupled channel
treatment of the $\pi \eta$ and $\pi \eta'$ systems.
The topological charge density mediates the coupling of 
the light-mass exotic state to the $\eta \pi$ and $\eta' \pi$ channels 
in our calculation.

\vspace{2.0cm}

{\bf Acknowledgements} \\

We thank K. Seth and W. Weise for helpful discussions.
This work was supported in part by the Alexander von Humboldt Foundation.

\end{document}